\documentclass[prb,reprint,floatfix,twoside,tbtags,showpacs]{revtex4-1}

\usepackage[utf8]{inputenc}
\usepackage{graphicx}
\usepackage{subfigure}
\usepackage{amsmath,amssymb}
\usepackage{bm}	% bold symbols in mathematics
\usepackage{xspace} % espaces apres nouvelles commandes
\usepackage[pdftex]{hyperref}

\newcommand{\etal}{\textit{et al}.\@\xspace}
\newcommand{\ie}{\textit{i.e.}\@\xspace}

\begin{document}

%<<< Title, author, ...
\title{Predicting dislocation climb: Classical modeling versus atomistic simulations}
\author{Emmanuel \surname{Clouet}}
\email{emmanuel.clouet@cea.fr}
\affiliation{Service de Recherches de Métallurgie Physique, CEA/Saclay,
91191 Gif-sur-Yvette, France}

\doi{10.1103/PhysRevB.84.092106}

\date{\today}
%>>>
\begin{abstract}
  The classical modeling of dislocation climb based 
  on a continuous description of vacancy diffusion
  is compared to recent atomistic simulations of dislocation climb
  in body-centered cubic iron under vacancy supersaturation
  [Phys. Rev. Lett. \textbf{105}, 095501 (2010)]. 
  A quantitative agreement is obtained, showing 
  the ability of the classical approach to describe 
  dislocation climb.
  The analytical model is then used to extrapolate
  dislocation climb velocities to lower dislocation densities,
  in the range corresponding to experiments. 
  This allows testing of the validity of the pure climb 
  creep model proposed by Kabir \etal
  [Phys. Rev. Lett. \textbf{105}, 095501 (2010)].
\end{abstract}
\maketitle

Dislocations can move out of their glide planes 
through the emission or absorption of point defects.
Such a mechanism, known as dislocation climb,
has been modeled for more than half a century now 
using a continuum description of matter and diffusion theory 
\cite{Mott1951,Weertman1955}.
One can thus find in most of the textbooks on dislocations 
\cite{Friedel1964,Hirth1982,Caillard2003,Nabarro1967,Indenbom1992}
analytical expressions which give the dislocation climbing velocity as a function 
of the applied stress, the temperature, the point defect supersaturation, etc.
Although this classical description has been shown to reasonably
explain experimental observations
\cite{Silcox1960,Seidman1966,Balluffi1969,Powell1975,Bonafos1998,Mompiou2008},
a quantitative validation by direct comparison to experiments 
is out of reach. 
Dislocation climb is indeed rarely the single mechanism producing plastic strain 
and one has usually to deal with a complex dislocation microstructure. 

On the other hand, atomistic simulations can be used to study the pure climb 
either of an isolated dislocation or of a well controlled dislocation microstructure.
Such simulations thus offer a natural way for a quantitative validation 
of the classical approach.
In this Brief Report, we compare predictions of the classical dislocation climb model
\cite{Friedel1964,Hirth1982,Caillard2003,Nabarro1967,Indenbom1992}
to the results of atomistic simulations published by Kabir \etal in Ref.~\onlinecite{Kabir2010}.

Kabir \etal\cite{Kabir2010} performed atomistic simulations 
to study the climb of a mixed dislocation in body-centered cubic iron.
Their simulations are based on a kinetic Monte Carlo algorithm which reproduces
the diffusion of vacancies and their annihilation on a dislocation jog.
The influence of the dislocation on the vacancy migration barriers 
is fully taken into account thanks to an empirical potential 
which allows them to search for the minimum energy path joining all neighboring 
vacancy stable positions (nudged elastic band method).
In these simulations, the dislocations can be considered as being saturated
with jogs because of the small dislocation length, 
and the climb is driven by a high vacancy supersaturation.

When pipe diffusion is fast enough and a high concentration of jogs is present on the dislocation,
one can assume that vacancies are at equilibrium all along the dislocation line
\cite{Friedel1964,Hirth1982,Caillard2003}.
Diffusion theory then predicts
\cite{Friedel1964,Hirth1982,Caillard2003,Mordehai2008,Bako2011}
that an infinite straight dislocation of Burgers vector $\mathbf{b}$ and character $\theta$
climbs at a velocity
\begin{equation*}
	v_{\rm cl} =  \eta 
	\frac{ D_{\rm V} }{ b \left| \sin{(\theta)} \right| 
		\ln{\left(R_{\infty}/r_{\rm c}\right)} }
	\left| C_{\rm V}^{\rm eq} - C_{\rm V}^{\infty} \right|.
\end{equation*}
The geometric factor $\eta=2\pi$ for an isolated dislocation.
$D_{\rm V}$ is the vacancy diffusion coefficient in a dislocation-free crystal, 
$C_{\rm V}^{\rm eq}$ the concentration of vacancies in equilibrium 
with the dislocation at a distance $r_{\rm c}\sim b$ from the line,
and $C_{\rm V}^{\infty}$ the average vacancy concentration imposed 
at a distance $R_{\infty}$.
This should correspond to half the average distance between dislocations,
\ie $R_{\infty}=1/(2\sqrt{\rho_{\rm D}})$ if $\rho_{\rm D}$ is the dislocation
density.
For a high vacancy supersaturation,
like in the atomistic simulations of Kabir \etal\cite{Kabir2010},
$C_{\rm V}^{\rm eq} \ll C_{\rm V}^{\infty}$
and then
\begin{equation}
	v_{\rm cl} = - \eta \frac{ D_{\rm V} }
	{ b \left| \sin{(\theta)} \right| \ln{\left(2r_{\rm c}\sqrt{\rho_{\rm D}}\right)} }
	C_{\rm V}^{\infty} .
	\label{eq:climb_velocity}
\end{equation}

\begin{figure}[tb]
	\begin{center}
		\includegraphics[width=0.8\linewidth]{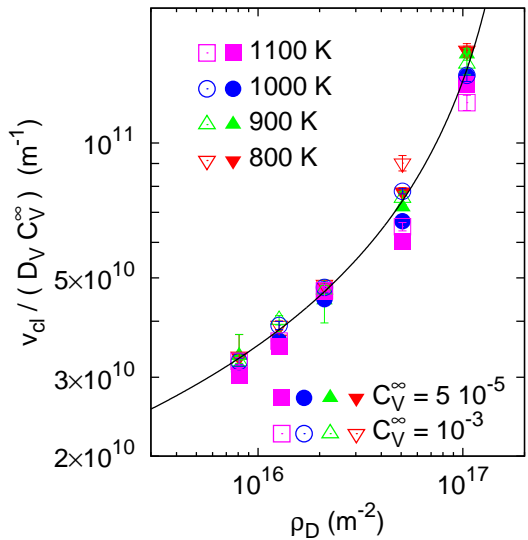}
	\end{center}
	\caption{Variation with the dislocation density $\rho_{\rm D}$
	of the climb velocity $v_{\rm cl}$ normalized by $D_{\rm V} C_{\rm V}^{\infty}$
	for different temperatures and vacancy supersaturations.
	Symbols corresponds to atomistic simulations \cite{Kabir2010}
	and lines to Eq.~(\ref{eq:climb_velocity}).}
	\label{fig:climb_velocity}
\end{figure}

\begin{figure*}[tb]
  \begin{center}
    \includegraphics[width=0.3\linewidth]{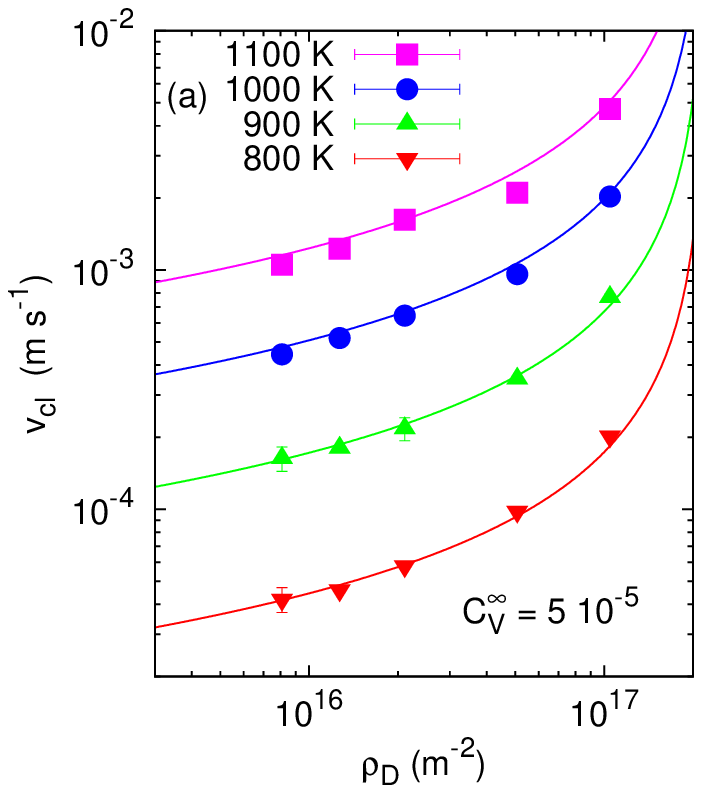}
    \hfill
    \includegraphics[width=0.3\linewidth]{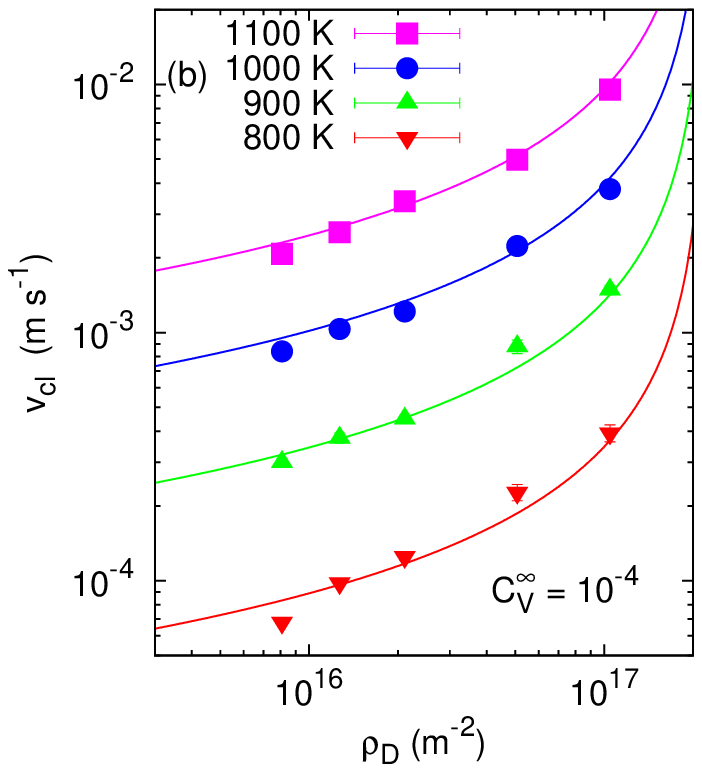}
    \hfill
    \includegraphics[width=0.3\linewidth]{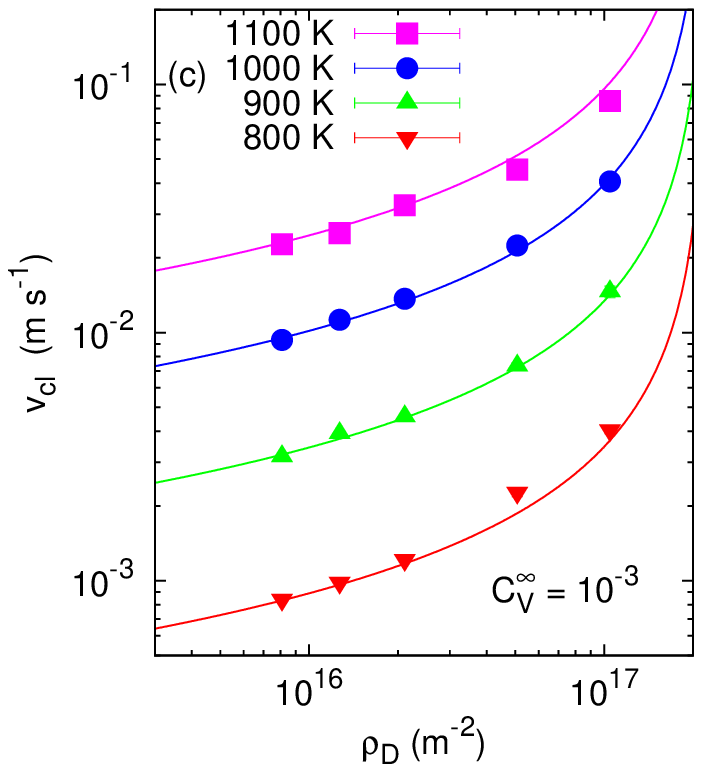}
  \end{center}
	\caption{Variation with the dislocation density $\rho_{\rm D}$
	of the climb velocity $v_{\rm cl}$ 
	for different temperatures and for a vacancy concentration
	(a) $C_{\rm V}^{\infty} = 5\,10^{-5}$,
	(b) $C_{\rm V}^{\infty} = 10^{-4}$,
	and (c) $C_{\rm V}^{\infty} = 10^{-3}$.
	Symbols corresponds to atomistic simulations \cite{Kabir2010}
	and lines to Eq.~(\ref{eq:climb_velocity}).}
	\label{fig:climb_velocity2}
\end{figure*}

Equation (\ref{eq:climb_velocity}) predicts that the climb velocity, 
once normalized by $D_{\rm V} C_{\rm V}^{\infty}$,
should only depend on $\rho_{\rm D}$.
This agrees with the atomistic simulations of Ref.~\onlinecite{Kabir2010}:
Results  for all temperatures and vacancy supersaturations
are well reproduced by Eq.~(\ref{eq:climb_velocity}).
(Figs.~\ref{fig:climb_velocity} and \ref{fig:climb_velocity2}).
The two parameters $\eta$ and $r_{\rm c}$ appearing in this equation
were used as fitting parameters, and the best quantitative agreement with 
atomistic simulations was obtained for the values $\eta=12.8$ and $r_{\rm c}=4.3b$.

The geometric factor obtained from this fit
($\eta \sim 4 \pi$) is close to its $2\pi$ theoretical value.
The slight difference may come from the fact that 
a periodic array of dislocation dipoles has been modeled in 
the atomistic simulations of Kabir \etal \cite{Kabir2010}, 
whereas an isolated dislocation is assumed in Eq.~(\ref{eq:climb_velocity}).
Burke and Nix \cite{Burke1978} also showed that the elastic interaction
between the vacancy and the dislocations, which is neglected in our modeling approach,
leads to a value for $\eta$ slightly higher than $2\pi$, 
thus in agreement with what we obtained.

The conventions are usually to take the capture radius $r_{\rm c}$
equal to the norm of the Burgers vector.
Our fit of Eq.~(\ref{eq:climb_velocity}) leads to $r_{\rm c}=4.3\ b$,
which actually agrees with the definition of the core region
($r<4\ b$) that has been deduced from previous atomistic calculations \cite{Lau2009}
on the same model system.

One therefore sees that the two parameters $\eta$ and $r_{\rm c}$ 
are physical parameters whose precise values are close to theoretical ones. 
It is also important to notice that both parameters do not theoretically depend
on the dislocation density, the vacancy supersaturation, the applied stress, nor the temperature. 
This was checked and all atomistic simulations of Ref. \onlinecite{Kabir2010} could be reproduced 
with a single set of parameters.

This comparison with atomistic simulations shows 
that the classical modeling of dislocation climb leads to quantitative predictions.
Such a model, based on a continuous description of vacancy diffusion,
does not explicitly take into account all atomic details of the vacancy 
diffusion close to the dislocation.
It nevertheless manages to perfectly reproduce results of atomistic simulations.
Finally, all atomistic details on the vacancy migration are not so relevant
to model dislocation climb
and one can use classical mesoscopic approaches
\cite{Friedel1964,Hirth1982,Caillard2003,Nabarro1967,Indenbom1992,Mordehai2008,Bako2011}
leading to simple analytical expressions like Eq.~(\ref{eq:climb_velocity}).

One can now use this expression of dislocation climb mobility 
to compare predictions of different creep models with experimental data, 
in particular the creep model developed by Kabir \etal \cite{Kabir2010}.
The dislocation densities in the atomistic simulations\cite{Kabir2010} 
($\rho_{\rm D} \gtrsim 10^{16}$\,m$^{-2}$) are much higher than the ones
usually observed in creep experiments in iron \cite{Garofalo1961,Karashima1971}
($\rho_{\rm D} \lesssim 10^{12}$\,m$^{-2}$).
As a consequence, one cannot directly use the dislocation climb velocities
measured in atomistic simulations,
but one needs to extrapolate them to lower dislocation densities.
Kabir \etal \cite{Kabir2010} used power laws to perform such an extrapolation
and concluded to the agreement of their model 
with creep experiments.
As such power laws do not rely on any physical ground, 
it is worth checking if the same nice agreement
can be obtained when the classical modeling of dislocation 
climb leading to Eq.~(\ref{eq:climb_velocity}) is used 
for this extrapolation.

In the creep model, one usually uses a power law
$v_{\rm cl} \propto {\rho_{\rm D}}^m$ to reproduce
the variations of the climbing velocity with the dislocation density.
Equation (\ref{eq:climb_velocity}) leads then to an exponent $m$ 
which depends on the density $\rho_{\rm D}$
\begin{equation}
  m = \frac{\partial \ln{(v_{\rm cl})}}{\partial \ln{(\rho_{\rm D})}}
  = \frac{-1}{2 \ln{\left(2r_{\rm c}\sqrt{\rho_{\rm D}}\right)}} .
  \label{eq:m_exponent}
\end{equation}
This exponent can now be used in the creep model proposed by Kabir \etal \cite{Kabir2010}.
This model assumes that all the plastic strain in creep is produced 
by climbing dislocations. 
Orowan law gives then the creep rate
$\dot{\varepsilon} = \rho_{\rm D} b v_{\rm cl}$.
The dislocation density is fixed by the Taylor relation 
($\rho_{\rm D} = (\sigma / \alpha G b )^2$ where $\alpha \sim 0.4$
is an empirical constant\cite{Kabir2010} and $G$ is the shear modulus).
The vacancy supersaturation, which varies linearly both 
with the applied stress $\sigma$ and the dislocation velocity,
is given by Eq.~(1) in Ref.~\onlinecite{Kabir2010}.
%($C_{\rm V}^{\infty} \propto \sigma v_{\rm cl}$).
Equation (\ref{eq:m_exponent}) combined with these assumptions leads then to 
a steady-state creep rate whose stress and temperature dependences
can be reproduced by the power law  
$\dot{\varepsilon}=\mathcal{A}\sigma^n\exp{(-Q/kT)}$ 
with a stress-dependent exponent $n=3+4m$: 
\begin{equation}
	n(\sigma)  
	= 3 - \frac{2}{\ln{\left( 2 r_{\rm c} \sigma / \alpha G b \right)}} .
	\label{eq:exponent}
\end{equation}

Equation (\ref{eq:exponent}) allows applying the creep model of Kabir \etal
to a stress range much lower than in atomistic simulations.
Stress in creep experiments in iron usually does not exceed 100\,MPa
\cite{Garofalo1961,Karashima1971,Karashima1966,Cadek1968,Cadek1969,Davies1973}.
Equation (\ref{eq:exponent}) then predicts an exponent $n$ smaller than 3.5,
thus far from the values higher than 6 found experimentally 
\cite{Cadek1969,Davies1973}.
One therefore sees that the creep model proposed by Kabir \etal
cannot explain experimental stress exponents measured in iron.
It naturally leads at low stress to an exponent close to 3,
like any other creep model based on pure climb \cite{Weertman1975}.
The high value for $n$ obtained by Kabir \etal \cite{Kabir2010}
corresponds to the much higher dislocation density
of their atomistic simulations
and cannot be directly compared to these experimental values.

One probably needs to consider both dislocation glide and climb
to obtain a stress exponent close to the experimental one.
A creep model based on dislocation pure climb is only valid
when dislocations cannot glide because of some constraints,
crystallographic constraints for instance 
as in hcp metals \cite{Edelin1973a,LeHazif1973}
or quasicrystals \cite{Mompiou2008,Mompiou2008a}.
Creep in alpha iron is far from this ideal case
as there is nothing preventing the dislocations from gliding.
%Kabir \etal introduce dislocation glide in their creep model\cite{Kabir2010}
%so as to produce a vacancy supersaturation
%but neglect then its contribution to the strain.
As pointed out by Weertman in his review paper \cite{Weertman1968},
``almost all of the creep strain is produced by glide motion 
of dislocations''.
%but [\ldots] the rate controlling process is
%the climb motion of dislocations''.
Although dislocation climb is the rate limiting process,
dislocation glide strongly affects creep and cannot be ignored.
Weertman \cite{Weertman1957,Weertman1968,Weertman1975} showed for instance 
that the consideration of glide in a creep model
makes the stress exponent increase from $n=3$ to $n=4.5$. 

Another discrepancy between the creep model of Kabir \etal \cite{Kabir2010}
and experiments \cite{Karashima1966,Cadek1968,Cadek1969} in iron
comes from the temperature dependence of the stress exponent $n$ 
and the stress dependence of the creep activation energy $Q$.
According to Eq.~(\ref{eq:exponent}), $n$ does not depend directly 
on the temperature.
A temperature dependence may only arise through a variation 
of the capture radius $r_{\rm c}$ with the temperature 
\footnote{The temperature dependence of the shear modulus $G$ is taken into account
by normalizing the applied stress $\sigma$ with $G$ when processing results 
of creep experiments. 
A temperature dependence of the parameter $\alpha$ entering Taylor relation 
is also possible but the modeling of such a variation will require 
a more complex approach than the simple creep model used here.}. 
As shown above, a single value of $r_{\rm c}$ could be used to reproduce
all the dislocation climbing velocities obtained by atomistic simulations
in the temperature range 800 -- 1100\,K.
As a consequence, this creep model does not lead
to any temperature dependence of the stress exponent $n$.
The creep rate is obtained from Orowan law 
$\dot{\varepsilon}=\rho_{\rm D}bv_{\rm cl}$
where the dislocation density is deduced from Taylor relation
and the climbing velocity is given by Eq.~(\ref{eq:climb_velocity}).
The activation energy $Q$ for creep is then the activation energy 
for vacancy diffusion whatever the applied stress.
The dependences found by Kabir \etal (Fig.~4 in Ref.~\onlinecite{Kabir2010})
are artifacts
caused by a fit  of the climbing velocity 
at high densities using a simple power law 
which neglects the logarithm appearing 
in Eq.~(\ref{eq:climb_velocity}).
To explain the dependence observed experimentally \cite{Karashima1966,Cadek1968,Cadek1969}, 
one has therefore to rely on mechanisms which are not considered
in the creep model of Kabir \etal.

Finally, it is worth pointing that the stress  
enters in the creep model of Kabir \etal \cite{Kabir2010} only through the control
of the dislocation density and of the vacancy supersaturation. 
As the climbing velocity does not depend then on the dislocation orientation,
all dislocations, whatever their orientations, are climbing at the same velocity.
If there is no specific climbing direction being enhanced by the stress,
no average macroscopic strain can develop and there will be no creep.

In summary, the comparison with atomistic simulations shows
that a classical approach at a mesoscopic scale
manages to quantitatively describe dislocation climb. 
Such an approach not only allows rationalizing results of atomistic simulations,
but it is also necessary to extrapolate them in a range of dislocation densities 
corresponding to experiments.
Thanks to such an extrapolation based on a physical sound model, 
a fair test of the validity of the creep model proposed by Kabir \etal \cite{Kabir2010}
can be made, thus showing its inability to reproduce experimental data 
on creep in iron.

\bibliography{clouet2011}
\bibliographystyle{apsrev4-1}

\end{document}